\documentclass[iop,tighten,twocolumn]{aastex631}

\usepackage{graphics,graphicx}
\usepackage{amsmath, amssymb}
\usepackage{multirow}
\usepackage{color}
\usepackage{verbatim}
\usepackage{hyperref}
\usepackage[normalem]{ulem}  
\usepackage{ulem}
\usepackage{color}
\usepackage{cancel}
\usepackage{mathtools}
\usepackage{epstopdf}

\renewcommand{\thefootnote}{\#\arabic{footnote}}

\renewcommand\sout{\bgroup\color{blue} \ULdepth=-.5ex \ULset}
\def\slashchar#1{\setbox0=\hbox{$#1$}  
\dimen0=\wd0     
\setbox1=\hbox{/} \dimen1=\wd1  
\ifdim\dimen0>\dimen1   
\rlap{\hbox to \dimen0{\hfil/\hfil}} 
#1     
\else     
\rlap{\hbox to \dimen1{\hfil$#1$\hfil}} 
/      
\fi}

\newcommand{\dd}{\mathrm{d}}

\newcommand{\eps}{\epsilon}

\begin{document}

\title{Curvature of the energy per particle in neutron stars}

\author{Micha\l{} Marczenko$^\dagger$}
\email{$^\dagger$michal.marczenko@uwr.edu.pl}
\affiliation{Incubator of Scientific Excellence - Centre for Simulations of Superdense Fluids, University of Wroc\l{}aw, plac Maksa Borna 9, PL-50204 Wroc\l{}aw, Poland}
\author{Krzysztof Redlich}
\affiliation{Institute of Theoretical Physics, University of Wroc\l{}aw, plac Maksa Borna 9, PL-50204 Wroc\l{}aw, Poland}
\affiliation{Polish Academy of Sciences PAN, Podwale 75, 
PL-50449 Wroc\l{}aw, Poland}
\author{Chihiro Sasaki}
\affiliation{Institute of Theoretical Physics, University of Wroc\l{}aw, plac Maksa Borna 9, PL-50204 Wroc\l{}aw, Poland}
\affiliation{International Institute for Sustainability with Knotted Chiral Meta Matter (WPI-SKCM$^2$), Hiroshima University, Higashi-Hiroshima, Hiroshima 739-8526, Japan}


\begin{abstract}

Neutron stars (NSs) serve as laboratories for probing strongly interacting matter at the most extreme densities. Their inner cores are expected to be dense enough to host deconfined quark matter. Utilizing state-of-the-art theoretical and multi-messenger constraints, we statistically determine the bulk properties of dense NS matter. We show that the speed of sound can be expressed in terms of the slope and curvature of the energy per particle. We demonstrate that the restoration of conformal symmetry requires changing the sign of the curvature of the bulk energy per particle as a function of energy density. Furthermore, we find that such a sign change is closely related to the peak in the speed of sound. We argue that the curvature of the energy per particle may serve as an approximate order parameter that signifies the onset of strongly coupled conformal matter in the NS core. 

\end{abstract}

\keywords{stars: neutron --- stars: interiors --- dense matter --- equation of state}


\section{Introduction}

The main extraterrestrial laboratories of dense nuclear matter are neutron stars (NSs), which can host matter up to several times the saturation density ($n_{\rm sat} = 0.16~\rm fm^{-3}$). The key quantity that determines the observed properties of these objects is the equation of state (EoS). One of the goals of multi-messenger astrophysics is to determine the EoS of dense nuclear matter. It is to be expected that NSs may contain exotic matter such as exotic hadrons, hadronic resonances, or even quark matter in their interiors. The appearance of exotic matter is usually associated with a strong first-order phase transition (FOPT), but other possibilities, e.g. a local peak in the speed of sound, are also not excluded. Such non-monotonic behaviors of the EoS are reflected in the speed of sound, which provides valuable insights into the microscopic description of dense matter
\begin{equation}\label{eq:cs2}
c_s^2 \equiv \frac{\dd p}{\dd \eps} = \frac{n}{\mu}\frac{\dd \mu}{\dd n} \textrm.
\end{equation}
Low-density nuclear matter ($n \lesssim 2n_{\rm sat}$) is faithfully described by chiral effective field theory ($\chi$EFT), which provided the first indications of violation of the conformal bound of $c_s^2=1/3$ in dense nuclear matter~\citep{Tews:2018kmu}. However, due to the non-perturbative nature of Quantum Chromodynamics (QCD) at low densities, first-principle calculations are only accessible through perturbative QCD (pQCD) methods at extremely high densities ($n \gtrsim 40 n_{\rm sat}$). Nevertheless, they confirm that the conformal limit is achieved asymptotically~\citep{Fraga:2013qra}. At intermediate densities, relevant for the phenomenology of NSs, great progress in constraining the EoS was achieved by systematic analyses of recent astrophysical observations of the massive pulsar PSRJ0740+6620~\citep{Cromartie:2019kug, Fonseca:2021wxt, Miller:2021qha, Riley:2021pdl} and PSR J0030+0451~\citep{Miller:2019cac} by the NICER collaboration, and the constraint from the GW170817 event~\citep{LIGOScientific:2018cki}, within parametric models of the EoS~(see, e.g.,~\cite{Hebeler:2013nza, Alford:2013aca, Kurkela:2014vha, Alford:2017qgh, Most:2018hfd, Li:2021sxb, Annala:2019puf, Somasundaram:2021clp, Annala:2017llu, Tews:2018iwm, Altiparmak:2022bke, Mroczek:2023zxo, Ecker:2022xxj, Jiang:2022tps, Brandes:2022nxa, Christian:2021uhd, Takatsy:2023xzf, Dietrich:2020efo, Landry:2020vaw, Al-Mamun:2020vzu, Miller:2021qha, Essick:2021kjb, Raaijmakers:2021uju, Huth:2021bsp, Capano:2019eae, Raaijmakers:2019dks, Essick:2019ldf, Han:2022rug}).

\begin{table*}[t!]
    \centering
    \begin{tabular}{|c||c|c|c|c|c|c|}\hline
                        &$\eps~\rm [GeV/fm^3]$ & $c_s^2$                            & $\gamma$                           & $\Delta$                                & $d_c$                                   & $\beta$                            \\\hline\hline
    $M_{1.4}$           & $0.406_{-0.024(0.066)}^{+0.027(0.105)}$ & $0.60^{+0.08(0.24)}_{-0.08(0.21)}$ & $3.94^{+0.43(1.62)}_{-0.39(1.15)}$ & $0.186^{+0.004(0.012)}_{-0.005(0.020)}$ & $0.48^{+0.07(0.23)}_{-0.06(0.17)}$      & $0.33^{+0.07(0.24)}_{-0.06(0.19)}$ \\\hline
    $c_{s, \rm peak}^2$ & $0.500^{+0.087(0.446)}_{-0.067(0.166)}$ & $0.86_{-0.07(0.26)}^{+0.06(0.13)}$ & $3.73_{-0.66(1.81)}^{+0.67(2.57)}$ & $ 0.11_{-0.05(0.20)}^{+0.04(0.11)}$     & $0.61_{-0.06(0.25)}^{+0.06(0.18)}$      & $0.46_{-0.07(0.27)}^{+0.06(0.20)}$ \\\hline
    $\beta=0$           & $0.678^{+0.091(0.382)}_{-0.076(0.221)}$ & $0.51_{-0.04(0.13)}^{+0.04(0.13)}$ & $1.49_{-0.04(0.13)}^{+0.04(0.13)}$ & $-0.01_{-0.03(0.13)}^{+0.03(0.11)}$     & $0.173_{-0.005(0.011)}^{+0.009(0.056)}$ & 0                                  \\\hline
    $M_{\rm TOV}$       & $1.073_{-0.070(0.202)}^{+0.071(0.267)}$ & $0.28^{+0.06(0.27)}_{-0.06(0.20)}$ & $0.80^{+0.13(0.59)}_{-0.15(0.53)}$ & $-0.02^{+0.03(0.10)}_{-0.03(0.11)}$     & $0.12^{+0.04(0.17)}_{-0.03(0.90)}$      & $-0.24^{+0.05(0.22)}_{-0.05(0.19)}$\\\hline
    pQCD              &  $\infty$ & $1/3$ & $1$ & $0$     & $0$      & $-1/6$\\\hline
    \end{tabular}
    \caption{Estimates of the selected properties of dense matter at the center of $1.4~M_\odot$ neutron star ($M_{1.4}$), at the peak of the speed of sound ($c_{s, \rm peak}^2$), changeover to conformal regime ($\beta=0$), maximally massive neutron star ($M_{\rm TOV}$), and conformal equation of state in the high-density limit (pQCD). Error bars are given at $1\sigma$ ($2\sigma$) confidence level.}
    \label{tab:properties}
\end{table*}

Recently, \cite{Fujimoto:2022ohj} proposed trace anomaly scaled by energy density as a new measure of conformality
\begin{equation}\label{eq:delta}
\Delta \equiv \frac{\eps - 3p}{3\eps} =  \frac{1}{3} - \frac{p}{\eps} \textrm.
\end{equation}
The vanishing $\Delta$ is a consequence of conformal invariance. The speed of sound in Eq. \eqref{eq:cs2} can be expressed in terms of $\Delta$ and its derivative as
\begin{equation}
    c_s^2 = \frac{1}{3} - \Delta - \eps \frac{\dd \Delta}{\dd \eps} \textrm.
\end{equation}
The swift increase of the speed of sound above its conformal value is related to the swift restoration of conformality~\citep{Fujimoto:2022ohj}. Interestingly, dense matter shows conformal behavior in the cores of the heaviest NSs at densities $\eps\simeq 1~\rm GeV/fm^3$~\citep{Fujimoto:2022ohj, Marczenko:2022jhl}. Another important quantity that is linked to the speed of sound and trace anomaly is the polytropic index
\begin{equation}
    \gamma \equiv \frac{\dd \log p}{\dd \log \eps} = \frac{\eps}{p} c_s^2 = \frac{c_s^2}{1/3 - \Delta}\rm.
\end{equation}
As the scale invariance becomes restored in QCD, $c_s^2 \rightarrow 1/3$ and $\Delta \rightarrow 0$, thus $\gamma \rightarrow 1$. 

Using the thermodynamic relations $\dd \eps = \mu \dd n$ and $ p = n^2 \dd (\eps/n)/ \dd n$, the speed of sound can be expressed in a different form as follows:
\begin{equation}
c_s^2 =\frac{1}{\mu}\frac{\dd p}{\dd n} = 2\frac{n}{\mu} \frac{\dd \eps/n}{\dd n} + \frac{n^2}{\mu}\frac{\dd^2 \eps/n}{\dd n^2} = \alpha + \beta\textrm,
\end{equation}
where
\begin{equation}
\begin{split}\label{eq:a_b}
    \alpha = 2\frac{c_s^2}{c_s^2+\gamma}= 2\frac{1/3 - \Delta}{4/3 - \Delta}\textrm,\;\;\;\;\;\;\;\;\beta  = c_s^2 - \alpha \textrm.
\end{split}
\end{equation}
Notably, $\alpha$ and $\beta$ are directly proportional to the slope and curvature of the bulk energy per particle $\eps/n$, respectively. At low densities, the conformal symmetry is broken, i.e., $\Delta \simeq 1/3$ and $c_s^2 \simeq 0$. Consequently, $\alpha \simeq 0$ and $\beta \simeq c_s^2$. On the other hand, at high densities, the system restores its conformal invariance, i.e., $c_s^2 \rightarrow 1/3$ and $\Delta \rightarrow 0$. In turn, $\alpha \rightarrow 1/2$ and $\beta\rightarrow -1/6$. Evidently, at low densities, $\beta$ is a non-negative increasing function of the energy density. Approaching the conformal limit, however, requires it to change the sign from positive to negative. From Eq.~\eqref{eq:a_b}, one sees that the curvature of the energy per particle vanishes for \mbox{$c_s^2 + \gamma =2$}. Moreover, the conformally broken ($\beta>0$) and restored ($\beta < 0$) phases can be characterized by $c_s^2 + \gamma > 2$ and $c_s^2 + \gamma < 2$, respectively. We note that $c_s^2 \in \left[0,1\right]$ and $\Delta \in \left[ -2/3,1/3\right]$, therefore $\alpha \in \left[0,1\right]$ and $\beta \in \left[-1,1\right]$.

In this work, we propose the curvature of the bulk energy per particle as an effective measure of conformality. We argue that the negative curvature signifies the onset of strongly-coupled conformal matter. We analyze its behavior statistically as well as in terms of a simple parametrized model of trace anomaly. We discuss possible implications for the phenomenology of NSs and relations to other important physical quantities.

\section{Methodology}

We construct an ensemble of EoSs based on the piecewise-linear speed-of-sound parametrization introduced by~\cite{Annala:2019puf}. The model has been already used in several other works~\citep{Annala:2021gom, Altiparmak:2022bke, Marczenko:2022jhl, Ecker:2022xxj, Annala:2021gom}. Here, we follow the prescription provided in~\cite{Altiparmak:2022bke}. At densities $n_B < 0.5~n_{\rm sat}$, we use the Baym-Pethick-Sutherland (BPS) EoS~\citep{Baym:1971pw}. In the range ($0.5 - 1.1)~n_{\rm sat}$, we use the monotrope EoS, $P = Kn_B^\Gamma$, where $\Gamma\in (1.77,3.23)$ is sampled randomly and $K$ is matched with the BPS EoS at $0.5~n_{\rm sat}$. At densities $n_B \gtrsim 40~n_{\rm sat}$, we use the pQCD results for the pressure, density, and speed of sound of cold quark matter in $\beta$-equilibrium~\citep{Fraga:2013qra}. The density and speed of sound can be calculated straightforwardly from the pressure. In this work, we use the pQCD results down to $\mu_{\rm pQCD}=2.6~\rm GeV$.

At densities $1.1~n_{\rm sat} \leq n_B \leq n(\mu_{\rm pQCD})$ we use the piecewise-linear parametrization of the speed of sound:
\begin{equation}
    c_s^2(\mu) = \frac{\left(\mu_{i+1} - \mu\right)c_{s,i}^2 + \left(\mu - \mu_i\right)c_{s,i+1}^2}{\mu_{i+1} - \mu_i} \textrm,
\end{equation}
where $\mu_i \leq \mu \leq \mu_{i+1}$. We generate $N$ pairs of $\mu_i$ and $c^2_{s,i}$, where $\mu_i \in [\mu(n_0),\mu_{\rm pQCD}]$ and $c_{s,i}^2 \in [0,1]$. The values of $\mu_1$ and $c^2_{s,1}$ are fixed by the values of the monotrope EoS at $n_0=1.1~n_{\rm sat}$, and $\mu_N = \mu_{\rm pQCD}$.

The net-baryon number density can be expressed as
\begin{equation}\label{eq:nb}
    n_B(\mu) = n_0 \exp{\int\limits_{\mu_0}^{\mu}\mathrm{d}\nu \;\frac{1}{\nu\;c_s^2(\nu)}} \textrm,
\end{equation}
where $n_0=1.1~n_{\rm sat}$ and $\mu_0 = \mu(n_0)$. Integrating Eq.~\eqref{eq:nb} gives the pressure:
\begin{equation}
    p(\mu) = p_0 + \int\limits_{\mu_0}^\mu \mathrm{d}\nu\; n_B(\nu) \textrm,
\end{equation}
where $p_0 = p(\mu(n_0))$. 

\begin{figure}
    \centering
    \includegraphics[width=.9\linewidth]{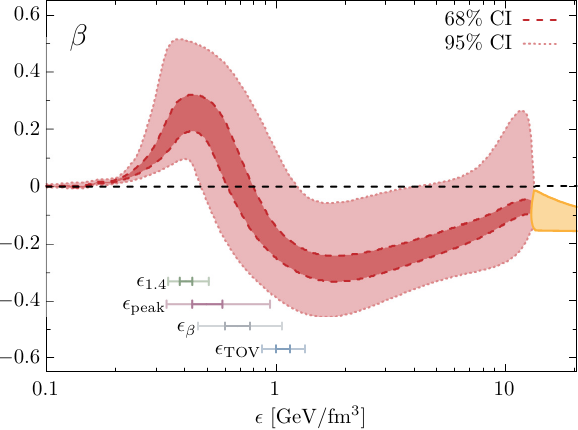}
    \caption{The curvature of the energy per particle, $ \beta$, as a function of energy density. Shown are results at $1\sigma$ ($68\%$) and $2\sigma$ ($95\textbf{}\%$) confidence intervals. Additionally, marked are $1\sigma$ and $2\sigma$ estimates of energy densities for centers of $1.4~M_\odot$ NSs (green), the peak of the speed of sound (purple), vanishing curvature of the energy per particle (grey), and centers of maximally massive NSs (blue). The orange band at high densities shows the pQCD constraint. Horizontal, dashed black line marks $ \beta = 0$.}
    \label{fig:beta}
\end{figure}

In addition to requiring consistency with the pQCD results at high densities, we impose observational astrophysical constraints. First, we utilize the GW170817 event measured by the LIGO/Virgo Collaboration (LVC). Viable EoSs should also result in a $1.4~M_\odot$ NS with tidal deformability of a $1.4~M_\odot$ NS, $\Lambda_{1.4} =190^{+390}_{-120}$~\citep{LIGOScientific:2018cki}. Second, we use the recently reported mass of the black widow pulsar PSR J0952-0607~\citep{Romani:2022jhd}, $M_{\rm TOV} = 2.35\pm0.17~M_\odot$, which is the heaviest detected neutron star. The pulsar is also among the fastest known rotating neutron stars and corrections are required to incorporate the equivalent non-rotating mass. As pointed out in~\citep{Brandes:2023hma}, using methodology from~\citep{Konstantinou:2022vkr}, the uncertainty in the correction is smaller than the uncertainty in the heavy-mass measurement itself. Therefore, we require the maximal mass to be at least the lower bound of the mass of PSR J0952-0607, i.e., $M_{\rm TOV} \geq 2.18~M_\odot$. In total, we analyzed a sample of $6\times10^5$ EoSs with $N=7$ that satisfy the pQCD boundary conditions, as well as the observational constraints from GW and pulsar measurements.

\section{Results}

\begin{figure}
    \centering
    \includegraphics[width=.9\linewidth]{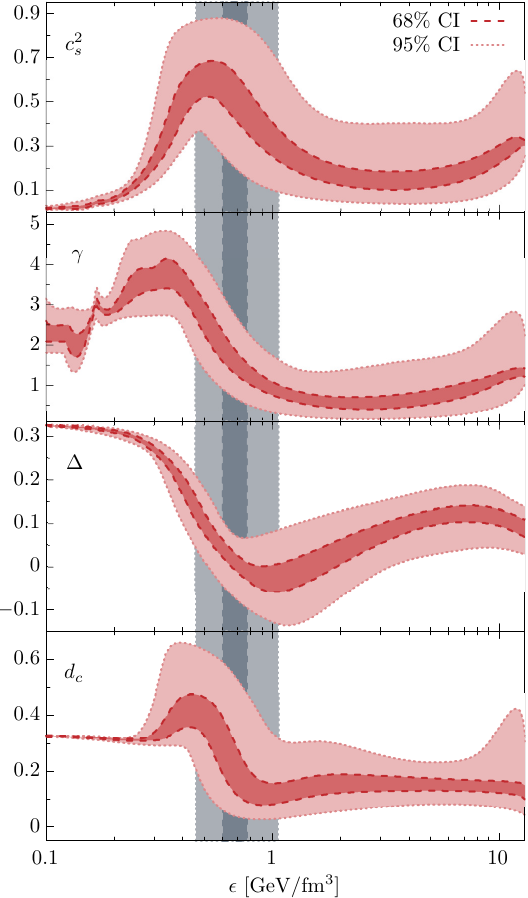}
    \caption{The speed of sound $c_s^2$, polytropic index $\gamma$, trace anomaly $\Delta$, measure of conformality $d_c$ as functions of energy density. Shown are results at $1\sigma$ ($68\%$) and $2\sigma$ ($95\%$) confidence intervals. The gray region shows $1\sigma$ and $2\sigma$ estimates of energy density for vanishing curvature of the energy per particle, i.e., $\beta=0$.}
    \label{fig:avg_panel}
\end{figure}

In Fig.~\ref{fig:beta}, we show the confidence intervals (CIs) of the curvature of the energy per particle $ \beta$ as a function of energy density. At low densities, $\beta$ increases and develops a maximum, similar to the speed of sound~\citep{Altiparmak:2022bke, Marczenko:2022jhl}. This is due to the low-density behavior being dominated by the swift increase in the speed of sound, while the trace anomaly changes only mildly~(cf.~Fig.~\ref{fig:avg_panel}). After $c_s^2$ develops a maximum, $\beta$ starts to decrease driven by the monotonic decrease of the trace anomaly $\Delta$. Consequently, $\beta$ becomes negative and reaches the pQCD constraint at high density with negative values. We note that the notable peak in $\beta$ at densities close to the imposed pQCD constraint is due to the parametrization method employed in this study, which allows for large variations of $c_s^2$ at chemical potential close to $\mu_{\rm pQCD}$ (see~\cite{Annala:2019puf, Altiparmak:2022bke, Jiang:2022tps}). This artifact should not affect our findings at densities below $\simeq \eps_{\rm TOV}$, which are relevant for the NSs phenomenology. 

The key result of this work is the determination of the energy density, $\eps_\beta$, for the vanishing curvature of the energy per particle $\beta$. We find the median $\eps_{\beta} = 0.678^{+0.091(0.382)}_{-0.076(0.221)}~\rm GeV/fm^3$ at $1\sigma(2\sigma)$ CI. We note that, in general, the current multi-messenger constraints do not forbid $\beta$ to change the sign multiple times. We find that this usually happens well above $\eps_{\rm TOV} = 1.073_{-0.070(0.202)}^{+0.071(0.267)}~\rm GeV/fm^3$ at $1\sigma$($2\sigma$) CI and, thus, is not relevant to the results reported in this work. In our sample, we find the position of the maximum of speed of sound at $\eps_{\rm peak} = 0.500^{+0.087(0.446)}_{-0.067(0.166)}~\rm GeV/fm^3$ at $1\sigma(2\sigma)$ CI. Remarkably, the curvature $\beta$ vanishes consistently at densities between $\eps_{\rm peak}$ and $\eps_{\rm TOV}$. We note that the peak in $c_s^2$ can be phenomenologically interpreted by connecting it to the phase boundary obtained in first-principles lattice QCD calculations and percolation threshold extracted from heavy-ion collision experiments~\citep{Marczenko:2022jhl}. The value of $\eps_{\rm peak}$ obtained in our work is also consistent with the critical percolation density obtained by~\cite{Marczenko:2022jhl}. Following the above discussion, one can conclude that the change in the sign of curvature of the energy per particle can be connected with the existence of the maximum in speed of sound. Therefore, the negative curvature of the energy per particle in the interior of NSs can be attributed to the change in medium composition.

\begin{figure}[t!]
    \centering
    \includegraphics[width=.9\linewidth]{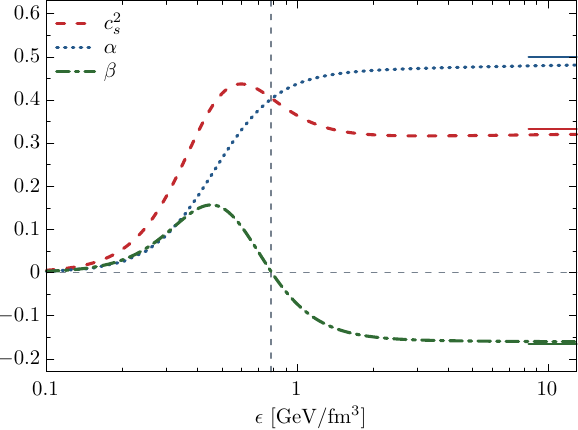}
    \caption{Curvature of the energy per particle $ \beta$ and its decomposition (see Eq.~\eqref{eq:a_b} for details) in the parametrized model from Eq.~\eqref{eq:delta_larry}. Solid lines indicate the limiting values in the conformally symmetric phase (see text). Gray, dashed lines mark $ \beta = 0$.}
    \label{fig:beta_larry}
\end{figure}

\begin{figure}
    \centering
    \includegraphics[width=.9\linewidth]{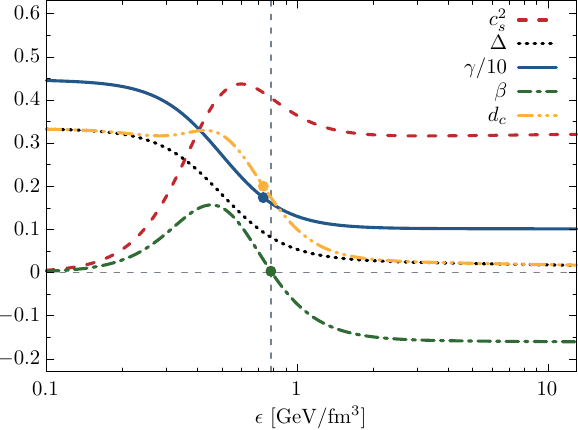}
    \caption{Comparison of physical quantities obtained in the parametrized model in Eq.~\eqref{eq:delta_larry}. Solid lines indicate the limiting values in the conformally symmetric phase (see text). Gray, dashed lines mark $\beta = 0$. Circles indicate criteria for identification of conformal matter, $\gamma=1.75$~\citep{Annala:2019puf}, $d_c=0.2$~\citep{Annala:2023cwx}, $\beta=0$ (this work).}
    \label{fig:comparison_larry}
\end{figure}

To qualitatively understand the structure of $ \beta$, we consider a minimal parametrization of monotonically decreasing $\Delta$ that fits the observational data~\citep{Fujimoto:2022ohj}
\begin{equation}\label{eq:delta_larry}
    \Delta\left(\eps\right) = \frac{1}{3} - \frac{1}{3}\frac{1}{e^{-\kappa\left(\eta-\eta_c\right)}+1}\left(1-\frac{A}{B+\eta^2}\right)\textrm,
\end{equation}
where $\kappa = 3.45$, $\eta_c=1.2$, $A=2$, $B=20$, $\eta=\log\eps/\eps_0$, and $\eps_0 = 0.15~\rm GeV/fm^3$. We note that $\Delta \geq 0$ in this model. In Fig.~\ref{fig:beta_larry}, we show $ \beta$ and its decomposition (see Eq.~\eqref{eq:a_b}) using the model introduced in Eq.~\eqref{eq:delta_larry}. At small densities, $ \alpha < c_s^2$ and $\beta$ is positive, owing to increasing $c_s^2$ and mildly changing $\Delta$. The most interesting is the behavior around $\eps = 0.4-0.9~\rm GeV/fm^3$. The speed of sound develops a peak at a value above the conformal value of $1/3$. Because $\Delta$ in Eq.~\eqref{eq:delta_larry} is a monotonic function, $ \alpha$ increases smoothly from $0$ to $1/2$. Eventually, the speed of sound develops a maximum, signaling swift restoration of conformality~\citep{Fujimoto:2022ohj}. Consequently, $ \alpha > c_s^2$, and $ \beta$ becomes negative at $\eps \approx 0.8 \rm ~ GeV/fm^3$. From the decomposition of $\beta$, it is clear that negative curvature of energy per particle signals the restoration of conformal symmetry. At high densities, the speed of sound has to converge to the conformal value $1/3$. Therefore, our interpretation is valid regardless of the height of the peak in $c_s^2$.

We also indicate that the additional requirement of the positive-definiteness of $\Delta$ allows us to find constraints for $c_s^2$ and $\gamma$ at which the curvature of energy per particle vanishes. From Eq.~\eqref{eq:a_b}, the condition $\beta = 0$ implies the following relations:
\begin{align}\label{eq:cs2_gamma_0}
    c_{s,\beta}^2 + \gamma_\beta = 2
\end{align}
and
\begin{align}\label{eq:delta_beta_0}
    \Delta_{\beta} = \frac{4}{3}\frac{c_{s,\beta}^2-1/2}{c_{s,\beta}^2 - 2} = \frac{4}{3}\frac{\gamma_\beta - 3/2}{\gamma_\beta}\textrm.
\end{align}
Under the assumption that $\Delta \geq 0$, from Eq.~\eqref{eq:delta_beta_0}, one gets that $c_{s,\beta}^2 \leq 1/2$ and $\gamma_\beta \geq 3/2$. We confirm that indeed in the parametrized model $\beta=0$ is characterized by $c_{s,\beta}^2 \approx 0.4$ and $\gamma_{\beta} \approx 1.6$, which is consistent with the extracted inequalities (see Fig.~\ref{fig:comparison_larry}). We note that, in principle, violation of the above inequalities would imply $\Delta < 0$.

In Fig.~\ref{fig:avg_panel}, we compile several physical quantities as functions of energy density: the speed of sound $c_s^2$, polytropic index $\gamma$, trace anomaly $\Delta$, and a measure of conformality $d_c$. The last quantity was recently proposed as an effective measure of the restoration of conformal symmetry and combines the trace anomaly $\Delta$ and its logarithmic derivative with respect to energy density $\Delta'$ into a single quantity ~\citep{Annala:2023cwx}
\begin{equation}
d_c = \sqrt{\Delta^2 + \left(\Delta'\right)^2}\label{eq:dc} \textrm.
\end{equation}
At low density $d_c \simeq 1/3$, but as the conformal symmetry is restored, $d_c\rightarrow 0$.

Interestingly, all quantities shown in Fig.~\ref{fig:avg_panel} manifest non-trivial behavior within our estimate for the energy density of vanishing curvature of energy per particle. The extracted values of the parameters are listed in Table~\ref{tab:properties}. They are consistent with the threshold values adopted in the literature $\gamma = 1.75$~\citep{Annala:2019puf} and $d_c=0.2$~\citep{Annala:2023cwx}. From the condition in Eq.~\eqref{eq:cs2_gamma_0} one may show that $d_{c,\beta} \gtrsim 0.16$. We also illustrate this in the parametrized model from Eq.~\eqref{eq:delta_larry} in Fig.~\ref{fig:comparison_larry} with a similar conclusion. We obtain $d_{c,\beta}\approx 0.17$, $c^2_{s,\beta} \approx 0.4$, and $\gamma_\beta \approx 1.6$. We also confirmed that the condition in Eq.~\eqref{eq:cs2_gamma_0} holds in the ensemble of EoSs considered in this work. Our results signify the role of the curvature of the energy per particle in quantifying the onset of deconfined/conformal matter.
 
Lastly, we remark on the first-order phase transition (FOPT). In general, a FOPT implies a plateau of constant pressure between two values of energy density, $\eps_{\rm L}$ and $\eps_{\rm R}$, i.e., $p(\eps_{\rm L} \leq \eps \leq \eps_{\rm R}) = p_{\rm T} = \it const$. In turn, the speed of sound vanishes $c_s^2(\eps_{\rm L} \leq \eps \leq \eps_{\rm R})=0$. On the other hand, the trace anomaly $\Delta$ increases from $\eps_{\rm L}$ to $\eps_{\rm R}$, i.e., $\Delta\left(\eps_{\rm R}\right) > \Delta\left(\eps_{\rm L}\right)$. Consequently, FOPT acts against the restoration of conformal symmetry. However, the curvature of the energy per particle $\beta$ becomes negative. Thus, the sign change of $\beta$ can potentially be also linked to the FOPT transition in dense matter. Nevertheless, having an array of various physical quantities at our disposal allows us to discriminate this scenario from a continuous restoration of conformality. We also note that at present there is very limited evidence favoring strong FOPT (see, e.g., studies based on Bayesian inference~\citep{Brandes:2023hma} and recent lattice QCD results featuring a peak in $c_s^2$ at vanishing temperature and finite isospin chemical potential~\citep{Brandt:2022hwy}). Certainly, new data from the next LIGO observation run (O5), as well as the third generation of gravitational-wave detectors (G3), will allow us to determine if there are discontinuities such as a FOPT in the cores of NSs.

\section{Conclusions}

We have statistically determined the bulk properties of the neutron star (NS) equation of state (EoS) in view of current multi-messenger constraints. We provided new arguments that dense matter in the cores of the most massive NSs may be almost conformal. We proposed a new criterion of conformality, the curvature of energy per particle. This quantity is positive at low densities and dominated by the swift increase in the speed of sound, which develops a peak, signaling fast approach to conformality~\citep{Fujimoto:2022ohj}. We demonstrated that the curvature of energy per particle must become negative as conformal symmetry is restored. We have shown that this changeover happens at energy densities between central energy densities of canonical $1.4~M_\odot$ and maximally massive NSs. Therefore, the most massive NSs may contain almost conformal matter. Moreover, we found that it is closely related to the position of the peak in the speed of sound, which strongly indicates a phase change in dense medium~\citep{Marczenko:2022jhl}. We also confirmed that our criterion is consistent with other criteria used in the literature.

It is desired to further explore our findings within effective models of the NS matter to determine microscopic characteristics of the QCD EoS related to the onset of the strongly coupled conformal matter.

\section*{Acknowledgements}
M.M. acknowledges fruitful discussions with Yuki Fujimoto and Oleksii Ivanytskyi. This work is supported partly by the Polish National Science Centre (NCN) under OPUS Grant No. 2022/45/B/ST2/01527 (K.R. and C.S.), and the program Excellence Initiative–Research University of the University of Wroc\l{}aw of the Ministry of Education and Science (M.M.). M.M. acknowledges the support of the European Union's Horizon 2020 research and innovation program under grant agreement No. 824093 (STRONG-2020). The work of C.S. was supported in part by the World Premier International Research Center Initiative (WPI) through MEXT, Japan. K.R. also acknowledges the support of the Polish Ministry of Science and Higher Education.

\bibliography{biblio}{}
\bibliographystyle{aasjournal}

\end{document}